\documentclass[final,5p,times,twocolumn,authoryear]{elsarticle}

\usepackage{amssymb}
\usepackage{array}

\usepackage{graphicx}
\usepackage{placeins}

\usepackage{xpatch}

\usepackage[font=normalfont,labelfont=bf]{caption}

\usepackage{etoolbox}
\patchcmd{\emailauthor}{(#2)}{}{}{}
\patchcmd{\urlauthor}{(#2)}{}{}{}

\journal{}

\makeatletter
\def\ps@pprintTitle{%
 \let\@oddhead\@empty
 \let\@evenhead\@empty
 \def\@oddfoot{}%
 \let\@evenfoot\@oddfoot}
\makeatother

\usepackage{xcolor}

\usepackage{hyperref}
\definecolor{links}{HTML}{01368e}

\makeatletter
\hypersetup{colorlinks=true}
\AtBeginDocument{\@ifpackageloaded{hyperref}
  {\def\@linkcolor{links}
   \def\@anchorcolor{links}
   \def\@citecolor{links}
   \def\@filecolor{links}
   \def\@urlcolor{links}
   \def\@menucolor{links}
   \def\@pagecolor{links}
\begingroup
  \@makeother\`%
  \@makeother\=%
  \edef\x{%
    \edef\noexpand\x{%
      \endgroup
      \noexpand\toks@{%
        \catcode 96=\noexpand\the\catcode`\noexpand\`\relax
        \catcode 61=\noexpand\the\catcode`\noexpand\=\relax
      }%
    }%
    \noexpand\x
  }%
\x
\@makeother\`
\@makeother\=
}{}}
\makeatother

\usepackage{filecontents}

\usepackage{natbib}

\begin{document}

\begin{frontmatter}



\title{TADPOLE Challenge: Prediction of Longitudinal Evolution in Alzheimer's Disease}


\ead{tadpole@cs.ucl.ac.uk}
\ead[url]{http://tadpole.grand-challenge.org}

\address[ucl]{Centre for Medical Image Computing, University College London, Gower Street, London, United Kingdom, WC1E 6BT}
\address[rotterdam]{Biomedical Imaging Group Rotterdam, Erasmus MC, Rotterdam, Netherlands, PO Box 2040 3000 CA}

% \address[mayo]{Department of Radiology, Mayo Clinic, 200 1st St SW, Rochester, United States, MN 55902}
% \address[portland]{Fariborz Maseeh Department of Mathematics and Statistics, Portland State University, 724 SW Harrison Street, Portland, United States, 97201}
\address[usc]{Laboratory of Neuro Imaging, Keck School of Medicine, University of Southern California, 2001 N Soto Street, Los Angeles, United States, CA 90032}
\address[ucsf]{Center for Imaging of Neurodegenerative Diseases, University of California San Francisco, 4150 Clement St. (114M), San Francisco, United States, CA 94121}
\address[drc]{Dementia Research Centre, University College London Institute of Neurology, London, United Kingdom, WC1N 3AX}

\address[amsterdam]{Department of Radiology and Nuclear Medicine, VU University Medical Centre, Amsterdam, Netherlands, 1081 HV}

% \cortext[core]{TADPOLE Core Team}

\author[ucl]{R\u{a}zvan V. Marinescu}
\author[ucl]{Neil P. Oxtoby}
\author[ucl]{Alexandra L. Young}
\author[rotterdam]{Esther E. Bron}
\author[usc]{Arthur W. Toga}
\author[ucsf]{Michael W. Weiner}
\author[drc,amsterdam]{Frederik Barkhof}
\author[drc]{Nick C. Fox}
% \author[mayo]{Clifford R. Jack Jr.}
% \author[portland]{Bruno M. Jedynak}

\author[rotterdam]{Stefan Klein}
\author[ucl]{Daniel C. Alexander}
\author{the EuroPOND Consortium, for the Alzheimer's Disease Neuroimaging Initiative}

% \address{University College London, Centre for Medical Image Computing, Gower Street, London, WC1E 6BT}

\begin{abstract}
The Alzheimer's Disease Prediction Of Longitudinal Evolution (TADPOLE) Challenge compares the performance of algorithms at predicting future evolution of individuals at risk of Alzheimer's disease. TADPOLE Challenge participants train their models and algorithms on historical data from the Alzheimer's Disease Neuroimaging Initiative (ADNI) study or any other datasets to which they have access. Participants are then required to make monthly forecasts over a period of 5 years from January 2018, of three key outcomes for ADNI-3 rollover participants: clinical diagnosis, Alzheimer's Disease Assessment Scale Cognitive Subdomain (ADAS-Cog13), and total volume of the ventricles. These individual forecasts are later compared with the corresponding future measurements in ADNI-3 (obtained after the TADPOLE submission deadline). The first submission phase of TADPOLE was open for prize-eligible submissions between 15 June and 15 November 2017. The submission system remains open via the website: \url{https://tadpole.grand-challenge.org}, although since 15 November 2017 submissions are not eligible for the first round of prizes. This paper describes the design of the TADPOLE Challenge.

\end{abstract}

\begin{keyword}
Alzheimer's disease \sep 
Disease prediction 
\sep Community Challenge 
\sep Biomarkers



\end{keyword}

\end{frontmatter}


\FloatBarrier
\section{Introduction}
\label{intro}

Alzheimer's disease (AD), and dementia in general, is a key challenge for 21st-century healthcare. The statistics are sobering \citep{winblad2016defeating}: in 2015, 47 million people worldwide suffer from dementia, of which AD is the most common cause; dementia costs \$818 billion worldwide, which is more than 1\% of the aggregaste global gross domestic product (GDP); AD might contribute to as many deaths as  does heart disease or cancer. There are no available treatments that can cure or even slow the progression of AD -- all clinical trials into putative treatments have failed to prove a disease-modifying effect. One key reason for these failures is the difficulty in identifying a group of patients at early stages of the disease, where treatments are most likely to be effective. 

While early and accurate diagnosis of dementia can be challenging, this can be aided by quantitative biomarker measurements taken from magnetic resonance imaging (MRI), positron emission tomography (PET), and cerebro-spinal fluid (CSF) samples extracted from lumbar puncture. It has been hypothesized for AD \citep{jack2010hypothetical,jack2013update,aisen2010clinical,frisoni2010clinical} that all these biomarkers become abnormal at different intervals before symptom onset, suggesting that together they can be used for accurate prediction of onset and overall disease progression in individuals. In particular, some of the early biomarkers become abnormal decades before symptom onset, and can thus facilitate early diagnosis. 

Several approaches for predicting AD-related target variables (e.g. clinical diagnosis, cognitive/imaging biomarkers) have been proposed which leverage multimodal biomarker data available in AD. Traditional longitudinal approaches based on statistical regression model the relationship of the target variables with other known variables. Examples include regression of the target variables against clinical diagnosis \citep{scahill2002mapping}, cognitive test scores \citep{yang2011quantifying, sabuncu2011dynamics}, rate of cognitive decline \citep{doody2010predicting}, and retrospectively staging subjects by time to conversion between diagnoses \citep{guerrero2016instantiated}. Another approach involves supervised machine learning techniques such as support vector machines, random forests, and artificial neural networks, which use pattern recognition to learn the relationship between the values of a set of predictors (biomarkers) and their labels (diagnoses). These approaches have been used to discriminate AD patients from cognitively normal individuals \citep{kloppel2008automatic, zhang2011multimodal}, and for discriminating at-risk individuals who convert to AD in a certain time frame from those who do not \citep{young2013accurate, mattila2011disease}. The emerging approach of disease progression modelling aims to reconstruct biomarker trajectories or other disease signatures across the disease progression timeline, without relying on clinical diagnoses or estimates of time to symptom onset. Examples include models built on a set of scalar biomarkers to produce discrete \citep{fonteijn2012event, young2014data} or continuous \citep{jedynak2012computational, donohue2014estimating, villemagne2013amyloid} biomarker trajectories; richer but less comprehensive models that leverage structure in data such as MR images \citep{durrleman2013toward, lorenzi2015disentangling, bilgel2016multivariate}; and models of disease mechanisms \citep{seeley2009neurodegenerative, zhou2012predicting, raj2012network, iturria2016early}.

These models have shown promise for predicting AD biomarker progression when using existing test data, but few have been tested on truly unseen \emph{future} data. Moreover, different investigators test these models on different datasets (including subsets of a single dataset) and use different processing pipelines. Community challenges have proved effective, in the medical image analysis field and beyond, for providing unbiased comparative evaluations of algorithms and tools designed for a particular task. Previous challenges that focussed on prediction of AD progression include the \emph{CADDementia challenge} \citep{bron2015standardized}, which aimed to predict clinical diagnosis from MRI scans. A similar challenge, the "\emph{International challenge for automated prediction of MCI from MRI data}" \citep{sarica2018machine} asked participants to predict diagnosis and conversion status from extracted MRI features of subjects from the ADNI study \citep{weiner2017recent}. Yet another challenge, The Alzheimer's Disease \emph{Big Data DREAM Challenge} \citep{allen2016crowdsourced}, asked participants to predict cognitive decline from genetic and MRI data. 

The Alzheimer's Disease Prediction Of Longitudinal Evolution (TADPOLE) Challenge aims to identify the data, features and approaches that are the most predictive of AD progression. In contrast to previous challenges, our motivation is to improve future clinical trials through identification of patients most likely to benefit from an effective treatment, i.e., those at early stages of disease who are likely to progress over the short-to-medium term (1-5 years). Identifying such subjects reliably helps cohort selection by focussing on groups that highlight positive treatment effects. The challenge thus focuses on forecasting three key features: clinical status, cognitive decline, and neurodegeneration (brain atrophy), over a five-year timescale. It uses “rollover” subjects from the ADNI study for whom a history of measurements is available, and who are expected to continue in the study, providing future measurements for testing. Since the test data does not exist at the time of forecast submissions, the challenge provides a completely unbiased basis for performance comparison. TADPOLE goes beyond previous challenges by drawing on a vast set of multimodal measurements from ADNI which support prediction of AD progression.

\FloatBarrier
\section{Competition Design}
\label{design}

\begin{figure*}
 \centering
 \includegraphics[width=0.7\textwidth]{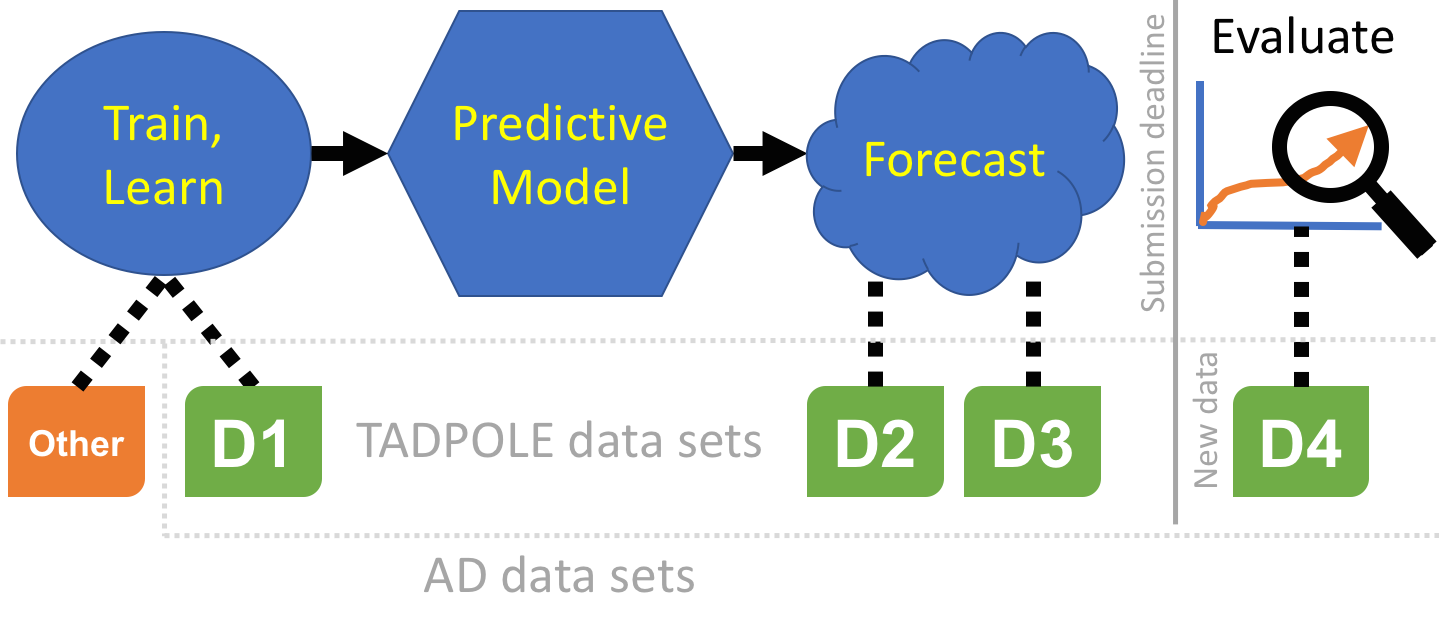}
 \caption{TADPOLE Challenge design. Participants are required to train a predictive model on a training dataset (D1 and/or others) and make forecasts for different datasets (D2, D3) by the submission deadline. Evaluation will be performed on a test dataset (D4) that is acquired after the submission deadline.}
 \label{fig:design}
 \end{figure*}

The aim of TADPOLE is to predict future outcome measurements of subjects at-risk of AD, enrolled in the ADNI study. A history of informative measurements from ADNI (imaging, psychology, demographics, genetics, etc.) from each individual is available to inform forecasts. TADPOLE participants are required to predict future measurements from these individuals and submit their predictions before a given submission deadline.  Evaluation of these forecasts occurs post-deadline, after the measurements have been acquired. A diagram of the TADPOLE flow is shown in Fig \ref{fig:design}.

\section{Forecasts}

Since we do not know the exact time of future data acquisitions for any given individual, TADPOLE challenge participants are required to make, for every individual, month-by-month forecasts of three key biomarkers: (1) clinical diagnosis which can be either cognitively normal (CN), mild cognitive impairment (MCI) or probable Alzheimer's disease (AD); (2) ADAS-Cog13 (ADAS13) score; and (3) ventricle volume (divided by intra-cranial volume). Evaluation is performed using forecasts at the months that correspond to data acquisition. TADPOLE forecasts are required to be probabilistic and some evaluation metrics will account for forecast probabilities provided by participants. Methods or algorithms that do not produce probabilistic estimates can still be used, by setting binary probabilities (zero or one) and default confidence intervals.

Participants are required to submit forecasts in a standardised format (see Table \ref{tab:subFormat}). For clinical status, relative likelihoods of each option (CN, MCI, and AD) for each individual should be provided. These are normalised at evaluation time; negative likelihoods are set to zero. For ADAS13 and ventricle volume, participants need to provide a best-guess value as well as a 50\% confidence interval for each individual. This 50\% confidence interval (as opposed to the more standard 95\%) was chosen to provide a more symmetric and less noisy evaluation of over- and under-estimation of the confidence interval, because similar sample sizes of data fall inside and outside the interval. 

\newcommand{\wi}{1.4cm}
\setlength\tabcolsep{3pt} % default value: 6pt

\begin{table*}
 \begin{tabular}{>{\centering\arraybackslash}m{0.6cm}  >{\centering\arraybackslash}m{1.2cm}  >{\centering\arraybackslash}m{1.2cm}  >{\centering\arraybackslash}m{1.7cm}  >{\centering\arraybackslash}m{1.7cm}  >{\centering\arraybackslash}m{1.7cm}  >{\centering\arraybackslash}m{0.9cm}  >{\centering\arraybackslash}m{1.2cm}  >{\centering\arraybackslash}m{1.2cm}  >{\centering\arraybackslash}m{1.4cm}  >{\centering\arraybackslash}m{1.5cm}  >{\centering\arraybackslash}m{1.5cm}}
  \textbf{RID} & \textbf{Forecast Month} & \textbf{Forecast Date} & \textbf{CN relative probability} & \textbf{MCI relative probability} & \textbf{AD relative probability} & \textbf{ADAS} & \textbf{ADAS 50\% CI lower} & \textbf{ADAS 50\% CI upper} & \textbf{Ventricles} & \textbf{Ventricles 50\% CI lower} & \textbf{Ventricles 50\% CI upper}\\
  \hline
  A & 1 & 2018-01 & 0 & 1 & 0 & 30 & 25 & 35 & 0.024 & 0.021 & 0.029\\
  B & 1 & 2018-01 & 3 & 2 & 0 & 25 & 21 & 26 & 0.023 & 0.021 & 0.025\\
  C & 1 & 2018-01 & 0.24 & 0.38 & 0.38 & 40 & 25 & 50 & 0.025 & 0.023 & 0.028\\
  
 \end{tabular}
  
 \caption{The format of the forecasts for three example subjects. }
 \label{tab:subFormat}
\end{table*}

% \FloatBarrier
\section{Data}

We provide participants with a standard ADNI-derived dataset (available via the Laboratory Of NeuroImaging: LONI) which they can use to train their algorithms, removing the need to pre-process the ADNI data themselves or merge different spreadsheets. However, participants are allowed to use a custom training set, by adding any other ADNI data or data from other studies. The software code used to generate the standard dataset is openly available in a Github repository\footnote{https://github.com/noxtoby/TADPOLE} and on the ADNI website, packaged with the standard dataset in the LONI ADNI database.

\subsection{ADNI data}

Data used in the preparation of this article were obtained from the Alzheimer's Disease Neuroimaging Initiative (ADNI) database (\url{adni.loni.usc.edu}). The ADNI was launched in 2003 by the National Institute on Aging (NIA), the National Institute of Biomedical Imaging and Bioengineering (NIBIB), the Food and Drug Administration (FDA), private pharmaceutical companies and non-profit organizations, as a \$60 million, 5-year public-private partnership. The initial goal of ADNI was to recruit 800 subjects but ADNI has been followed by ADNI-GO and ADNI-2. To date these three protocols have recruited over 1500 adults, ages 55 to 90, to participate in the research, consisting of cognitively normal older individuals, people with early or late MCI, and people with early AD. The general ADNI inclusion criteria has been described in \cite{petersen2010alzheimer}. 

The data we used from ADNI consists of: (1) CSF markers of amyloid-beta and tau deposition; (2) various imaging modalities such as magnetic resonance imaging (MRI), positron emission tomography (PET) using several tracers: Fluorodeoxyglucose (FDG, hypometabolism), AV45 (amyloid), AV1451 (tau) as well as diffusion tensor imaging (DTI); (3) cognitive assessments acquired in the presence of a clinical expert; (4) genetic information such as alipoprotein E4 (APOE4) status extracted from DNA samples; and (5) general demographic information. Extracted features from this data were merged together into a final spreadsheet and made available on the LONI ADNI website.

\subsection{Image pre-processing}

The imaging data has been pre-processed with standard ADNI pipelines.  For MRI scans, this included correction for gradient non-linearity, B1 non-uniformity correction and peak sharpening\footnote{see MRI analysis on ADNI website: \url{http://adni.loni.usc.edu/methods/mri-analysis/
mri-pre-processing}}. Meaningful regional features such as volume and cortical thickness were extracted using the Freesurfer cross-sectional and longitudinal pipelines \citep{reuter2012within}. Each PET image (FDG, AV45, AV1451), which consists of a series of dynamic frames, had its frames co-registered, averaged across the dynamic range, standardised with respect to the orientation and voxel size, and smoothed to produce a uniform resolution of 8mm full-width/half-max (FWHM)\footnote{see PET analysis on ADNI website: \url{http://adni.loni.usc.edu/methods/pet-analysis/pre-processing}}. Standardised uptake value ratio (SUVR) measures for relevant regions-of-interest were extracted (see \cite{jagust2010alzheimer}) after registering the PET images to corresponding MR images using the SPM5 software \citep{ashburner2009computational}. DTI scans were corrected for head motion and eddy-current distortion, skull-stripped, EPI-corrected, and finally aligned to the T1 scans using the pipeline from \cite{nir2013effectiveness}. Diffusion tensor summary measures were estimated based on the Eve white-matter atlas by \cite{oishi2009atlas}. 

% \FloatBarrier
\section{TADPOLE Datasets}
\label{datasets}

The TADPOLE Challenge involves three kinds of data sets: (a) a \emph{training data set}, which is a collection of measurements with associated outcomes that can be used to fit models or train algorithms; (b) a \emph{prediction data set}, which contains only baseline measurements (possibly longitudinal), without associated outcomes --- this is the data that algorithms, models, or experts use as input to make their forecasts of later patient status and outcome; and (c) \emph{a test data set}, which contains the patient outcomes against which we will evaluate forecasts --- in TADPOLE, this data did not exist at the time of submitting forecasts.

In order to evaluate the effect of different methodological choices, we prepared three “standard” data sets for training and prediction: 
\begin{itemize}
 \item \textbf{D1}: The TADPOLE \underline{\smash{standard training set}} draws on longitudinal data from the entire ADNI history. The data set contains a set of measurements for every individual that has provided data to ADNI in at least two separate visits (different dates) across three phases of the study: ADNI1, ADNI GO, and ADNI2. 
 \item \textbf{D2}: The TADPOLE \underline{\smash{longitudinal prediction set}} contains as much available data as we could gather from the ADNI rollover individuals for whom challenge participants are asked to provide forecasts. D2 includes all available time-points for these individuals. 
 \item \textbf{D3}: The TADPOLE \underline{\smash{cross-sectional prediction set}} contains a single (most recent) time point and a limited set of variables from each rollover individual in D2. Although we expect worse forecasts from this data set than D2, D3 represents the information typically available when selecting a cohort for a clinical trial. 
\end{itemize}

\begin{figure*}
 \centering
 \includegraphics[width=0.7\textwidth]{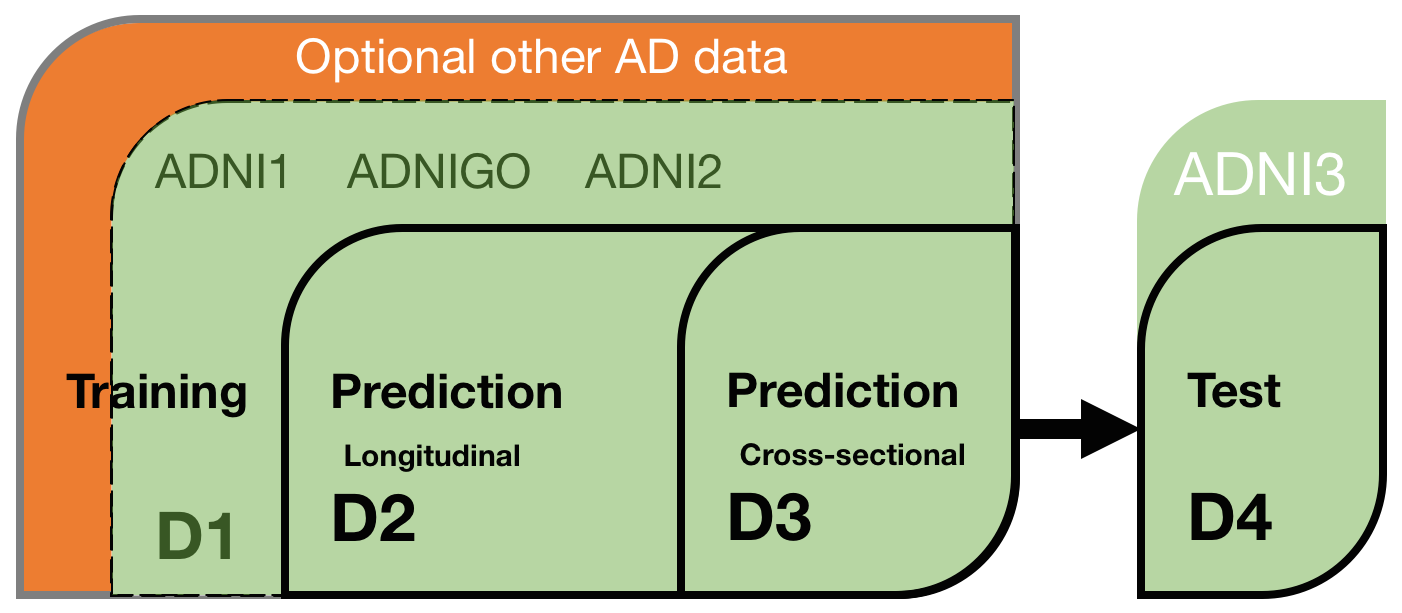}
 \caption{Venn diagram of the ADNI datasets for training (D1), longitudinal prediction (D2), cross-sectional prediction (D3) and the test set (D4). D3 is a subset of D2, which in turn is a subset of D1. Other non-ADNI data can also be used for training.}
 \label{fig:venn_diagram}
\end{figure*}

The forecasts will be evaluated on future data (D4 -- test set) from ADNI3 rollovers, acquired after the challenge submission deadline. In addition to the three standard datasets (D1, D2 and D3), challenge participants are allowed to use any other data sets that might serve as useful additional training data.  

Fig. \ref{fig:venn_diagram} shows a diagram highlighting the nested structure of datasets D1--D3. Table \ref{tab:biomk_data_available} shows the proportion of biomarker data available in each dataset. There are a considerable number of entries with missing data, especially for some biomarkers such as tau imaging (AV1451). We also estimated the expected number of subjects and available data for D4, using information from the ADNI3 procedures and using rollovers from previous ADNI studies (Table \ref{tab:biomk_data_available}, right-most column) -- See \ref{app:expectedD4} for more information on D4 estimates. Based on our estimates, we believe the size of D4 (around 330 subjects, 1 visit/subject) should be enough for a reliable evaluation of TADPOLE submissions.

\begin{table}
\centering
 \begin{tabular}{c | c | c c c c}
%   \multicolumn{6}{l}{}\\
 \multicolumn{2}{c|}{\textbf{Subject statistics}} & D1 & D2 & D3 & D4 \\
 \hline
 \multicolumn{2}{c|}{Nr. of subjects } & 1667 & 896 & 896 & \emph{330}\\
 \multicolumn{2}{c|}{Visits per subject }&  $\mathbin{{7.6}{\pm}{3.8}}$  & $\mathbin{{8.5}{\pm}{4.2}}$ & $\mathbin{{1.0}{\pm}{0.0}}$ & $\mathit{\mathbin{{1.0}{\pm}{0.0}}}$\\
 & CN & 31 & 38 & 45 & \emph{39} \\
 Diagnosis* (\%) & MCI & 56 & 57 & 39 & \emph{49} \\
 & AD & 13 & 5 & 16 & \emph{12} \\
%  \multicolumn{6}{c}{}\\
 \multicolumn{2}{l}{\textbf{Data availability**}}\\
 \hline
 \multicolumn{2}{c|}{Cognitive tests (\%) } & 70 & 68 & 84 & \emph{62} \\
 \multicolumn{2}{c|}{MRI (\%) } & 62 & 56 & 75 & \emph{69} \\
 \multicolumn{2}{c|}{FDG-PET (\%) } & 16 & 20 & 0 & \emph{20} \\
 \multicolumn{2}{c|}{AV45-PET (\%) } & 16 & 22 & 0 & \emph{19} \\
 \multicolumn{2}{c|}{AV1451-PET (\%) } & 0.7 & 1.1 & 0 & \emph{19} \\
 \multicolumn{2}{c|}{DTI (\%) } & 6 & 8 & 0 & \emph{15} \\
 \multicolumn{2}{c|}{CSF (\%) } & 18 & 19 & 0 & \emph{14} \\
 \end{tabular}
  
 \caption{Biomarker summary of TADPOLE datasets D1, D2 and D3. There is considerable amount of missing data in some biomarkers such as AV1451. Numbers for D4 are estimated based on ADNI3 procedures (see ADNI3 procedures manual) and rollovers from previous ADNI studies. (*) Diagnosis at baseline visit. (**) Percentage of all visits (across all subjects) that have measurements for desired biomarker.}
 \label{tab:biomk_data_available}
\end{table}

\section{Submissions}
\label{submissions}

There are two kinds of submissions that challenge participants can make. A simple entry requires a minimal forecast and a description of methods; it makes participants eligible for the prizes but not co-authorship on the scientific paper documenting the results. A simple entry can use any training data or prediction sets and forecast at least one of the target outcome variables (clinical status, ADAS13 score, or ventricle volume). A full entry entitles participants for consideration as a co-author on the publication documenting the results. Such a full entry requires a complete forecast for all three outcome variables on all subjects from the D2 prediction set, along with a description of the methods. Each individual participant is limited to a maximum of three submissions. This restriction has been introduced to avoid the risk of participants “tuning” their method on the test set by submitting multiple predictions for a range of algorithm settings. Although not required for a full entry, participants are strongly encouraged to submit predictions also for D3. 

Prizes are awarded to the best entries regardless of the choice of training sets (D1/custom) and prediction sets (D2/D3). However, the additional submissions support the key scientific aims of the challenge by allowing us to separate the influence of the choice of training data, post-processing pipelines, and modelling techniques or prediction algorithms. The target variables used for evaluation, in particular ventricle volume, will use the same post-processing pipeline as the standard data sets D1-D3.

Beyond the standard training dataset (D1), participants can include additional forecasts from "custom" (i.e. constructed by the participant) training data or custom post-processing of the raw data from subjects in the standard training set. The same applies to the prediction sets D2 and D3, which can be customised by the participants if desired, e.g. a prediction set with different features from the same individuals as in D2 and D3. Table \ref{tab:submissions} shows the twelve possible combinations of subject sets, processing and prediction sets, from which a full-entry submission must contain at least one of the first four (ID 1--4).

\begin{table}
\centering
 \begin{tabular}{c | c | c | c}
\textbf{ID} & \multicolumn{2}{c|}{\textbf{Training set}} & \textbf{Prediction set}\\
& Subject set & Post-processing & \\
\hline
1 & D1 & standard & D2\\
2 & D1 & custom & D2\\
3 & custom & standard & D2\\
4 & custom & custom & D2\\
5 & D1 & standard & D3\\
6 & D1 & custom & D3\\
7 & custom & standard & D3\\
8 & custom & custom & D3\\
9 & D1 & standard & custom\\
10 & D1 & custom & custom\\
11 & custom & standard & custom\\
12 & custom & custom & custom\\
  
\end{tabular}
\caption{Types of submissions that can be made by participants, for different types of training sets, prediction sets and post-processing pipelines.}
\label{tab:submissions}
\end{table}

% \FloatBarrier
\section{Forecast Evaluation}
\subsection{Clinical Status Prediction}

For evaluation of clinical status predictions, we will use similar metrics to those that proved effective in the CADDementia challenge \citep{bron2015standardized}: (i) the multiclass area under the receiver operating curve (mAUC); and (ii) the overall balanced classification accuracy (BCA). The mAUC is independent of the group sizes and gives an overall measure of classification ability that accounts for relative likelihoods assigned to each class. The simpler BCA is also independent of group sizes, but does not exploit the probabilistic nature of the forecasts. 

\subsubsection{Multiclass Area Under the Receiver Operating Characteristic (ROC) Curve}

The multiclass Area Under the ROC Curve (mAUC) is a simple generalisation of the area under the ROC curve applicable to problems with more than two classes \citep{hand2001simple}. The AUC $\hat{A}(c_i|c_j)$ for classification of a class $c_i$ against another class $c_j$, is:
\begin{equation}
\hat{A}(c_i|c_j)=\frac{S_i-n_i(n_i+1)/2}{n_i n_j}
\end{equation}
where $n_i$ and $n_j$ are the number of points belonging to classes $i$ and $j$, respectively; while $S_i$ is the sum of the ranks of the class $i$ test points after ranking all the class $i$ and $j$ data points in increasing likelihood of belonging to class $i$. We further define the average AUC for classes $i$ and $j$ as $\hat{A}(c_i,c_j)= 0.5(\hat{A}(c_i|c_j)+\hat{A}(c_j|c_i))$. The overall mAUC is then obtained by averaging $\hat{A}(c_i,c_j)$ over all pairs of classes:
\begin{equation}
 mAUC = \frac{2}{L(L-1)}\sum_{i=2}^L\sum_{j=1}^{i}\hat{A}(c_i,c_j)
\end{equation}
where $L$ is the number of classes. The class probabilities that go into the calculation of $S_i$ in the first equation are $p_{CN}$, $p_{MCI}$ and $p_{AD}$, which are derived from the likelihoods of each ADNI subject being assigned to each diagnostic class, by normalising to have unity sum.

\subsubsection{Balanced Classification Accuracy}

The Balanced Classification Accuracy (see \cite{brodersen2010balanced}) is an extension of the classification accuracy measure that accounts for the imbalance in the numbers of datapoints belonging to each class. However, the measure is not probabilistic, so in TADPOLE the data points need to be assigned a hard classification to the class (CN, MCI, or AD) with the highest likelihood. The balanced accuracy for class $i$ is then:
\begin{equation}
 BCA_i = \frac{1}{2}\left[\frac{TP}{TP+FN}+\frac{TN}{TN+FP}\right]
\end{equation}
where TP, FP, TN, FN represent the number of true positives, false positives, true negatives and false negatives for classification as class $i$. In this case, true positives are data points with true label $i$ and correctly classified as such, while the false negatives are the data points with true label $i$ and incorrectly classified to a different class $j \ne i$. True negatives and false positives are defined similarly. The overall BCA is given by the mean of all the balanced accuracies for every class. 

\subsection{Continuous Feature Predictions}

For ADAS13 and ventricle volume, we will use three metrics: mean absolute error (MAE), weighted error score (WES) and coverage probability accuracy (CPA). The MAE focuses purely on accuracy of the best-guess prediction ignoring the confidence interval, whereas the WES incorporates confidence estimates into the error score. The CPA provides an assessment of the accuracy of the confidence estimates, irrespective of the best-guess prediction accuracy.

\subsubsection{Mean Absolute Error}

The mean absolute error (MAE) is:
\begin{equation}
 MAE = \frac{1}{N}\sum_{i=1}^{N}\left|{\tilde{M}_i-M_i}\right|
\end{equation}
where $N$ is the number of data points (forecasts) evaluated, $M_i$ is the actual biomarker value in individual $i$ in future data, and $\tilde{M}_i$ is the participant's best prediction for $M_i$.

\subsubsection{Weighted Error Score}

The weighted error score is defined as:
\begin{equation}
 WES=\frac{\sum_{i=1}^{N}\tilde{C}_i\left|\tilde{M}_i-M_i\right|}{\sum_{i=1}^{N}\tilde{C}_i}
\end{equation}
where the weightings $\tilde{C}_i$ are the participant's relative confidences in their $\tilde{M}_i$. We estimate $\tilde{C}_i$ as the inverse of the width of the 50\% confidence interval of their biomarker estimate:
\begin{equation}
\tilde{C}_i=\left(C_+-C_-\right)^{-1}
\end{equation}
where $[C-, C+]$ is the confidence interval provided by the participant.

\subsubsection{Coverage Probability Accuracy}

The coverage probability accuracy is:
\begin{equation}
CPA = |ACP - NCP| 
\end{equation}
where $NCP$ is the nominal coverage probability, the target for the confidence intervals, and $ACP$ is the actual coverage probability, defined as the proportion of measurements that fall within the corresponding confidence interval. In TADPOLE, we set $NCP$ to be 0.5, which means that ideally only 50\% of the measurements would fall inside the confidence interval. The CPA can take values between 0 and 1, and lower scores are better.

\section{Prizes}
We are extremely grateful to Azheimer's Research UK, The Alzheimer's Society, and The Alzheimer's Association for sponsoring a prize fund of £30,000. At the time of first submission, we proposed six separate prizes, as outlined in Table \ref{tab:prizes}, but reserve the right to reallocate the prize money depending on the numbers of participants eligible for each prize. The first four are general categories (open to all challenge participants) and constitute one prize for the best forecast of each feature as well as one for overall best performance. The last two prizes are for two different student categories.

\begin{table}
\centering
 \begin{tabular}{>{\centering\arraybackslash}m{1.5cm}  c  >{\centering\arraybackslash}m{2cm}  >{\centering\arraybackslash}m{2cm}}
\textbf{Prize amount} & \textbf{Outcome measure} & \textbf{Performance Metric} & \textbf{Eligibility} \\
\hline
£5,000 & Clinical status & mAUC & all \\
£5,000 & ADAS13 & MAE & all\\
£5,000 & Ventricle volume & MAE & all\\
£5,000 & Overall best & Lowest sum of ranks* & all\\
£5,000 & Clinical status & mAUC & University teams\\
£5,000 & Clinical status & mAUC & High-school teams\\
\end{tabular}
\caption{Prize allocation scheme using funds from Azheimer's Research UK, The Alzheimer's Society and The Alzheimer's Association. There are 6 prizes awarded to different outcome measures, the last two of which are eligible only for university and high-school teams. (*) The overall best team will be the team that obtains the lowest sum of ranks in the clinical status, ADAS13 and ventricle volume categories. }
\label{tab:prizes}
\end{table}

\section{Discussion}

We have outlined the design of the TADPOLE Challenge, which aims to identify algorithms and features that can best predict the evolution of Alzheimer's disease. Challenge participants use historical data from ADNI in order to predict three key outcomes: clinical diagnosis, ADAS-Cog13 and ventricle volume. Determining which features and algorithms best predict AD evolution can aid refinement of cohorts and endpoint assessment for clinical trials, and can provide accurate prognostic information in clinical settings. 

The TADPOLE Challenge was designed to be transparent and accessible. To this end, all of our scripts are available in an open repository\footnote{TADPOLE repository: https://github.com/noxtoby/TADPOLE}. We also created a public forum\footnote{TADPOLE forum:  https://groups.google.com/forum/\#!forum/tadpolechallenge} where we answer participant questions. Finally, in order to enable participants to share algorithm performance results throughout the competition, we created a leaderboard system\footnote{Leaderboard: https://tadpole.grand-challenge.org/leaderboard/} that evaluates submissions on an existing test dataset and publishes the results live on our website.  

Going forward, we hope that by November 2018 sufficient data will be available from ADNI3 rollovers for a first meaningful evaluation of the forecasts. We plan to publish the results on the website in January 2019, and then submit a publication of the results soon after. However, we reserve the right to delay evaluation until sufficient data is available. At that time, we will also evaluate the impact and interest of the first phase of TADPOLE within the community, to guide decisions on whether to organise further submission and evaluation phases.

\FloatBarrier
\section{Acknowledgements}

% \address[mayo]{Department of Radiology, Mayo Clinic, 200 1st St SW, Rochester, United States, MN 55902}
% \address[portland]{Fariborz Maseeh Department of Mathematics and Statistics, Portland State University, 724 SW Harrison Street, Portland, United States, 97201}

TADPOLE Challenge has been organised by the European Progression Of Neurological Disease (EuroPOND) consortium, in collaboration with the ADNI. We thank all the participants and advisors, in particular Clifford R. Jack Jr., Mayo Clinic, Rochester, United States and Bruno M. Jedynak, Portland State University, Portland, United States for useful input and feedback.

The organisers are extremely grateful to Azheimer's Research UK, The Alzheimer's Society, and The Alzheimer's Association for sponsoring the challenge by providing the prize fund and providing invaluable advice into its construction and organisation. Similarly, we thank the ADNI leadership and members of our advisory board and other members of the EuroPOND consortium for their valuable advice and support.

RVM is supported by the EPSRC Centre For Doctoral Training in Medical Imaging with grant EP/L016478/1. NPO, FB, SK, and DCA are supported by EuroPOND, which is an EU Horizon 2020 project. ALY is currently supported by an EPSRC Doctoral Prize fellowship and was previously supported by EPSRC grant EP/J020990/01. DCA is supported by EPSRC grants J020990, M006093 and M020533. Data collection and sharing for this project was funded by the Alzheimer's Disease Neuroimaging Initiative (ADNI) (National Institutes of Health Grant U01 AG024904) and DOD ADNI (Department of Defense award number W81XWH-12-2-0012). FB is supported by the NIHR UCLH biomedical research centre and the AMYPAD project, which has received support from the EU-EFPIA Innovative Medicines Initiatives 2 Joint Undertaking (AMYPAD project, grant 115952). This project has received funding from the EU Horizon 2020 research and innovation programme under grant agreement No 666992.

\appendix

\section{Expected number of subjects and available data for D4}
\label{app:expectedD4}

We estimated the number of subjects and available data in D4 (Table \ref{tab:biomk_data_available}, last column) using information from the ADNI procedures manual and previous ADNI rollovers. For estimating the total number of subjects (first row) expected in D4, we computed the dropout rate (0.36) based on ADNI1 rollovers to ADNI2, then multiplied it by the total number of subjects in D2 (896). For estimating the proportions of each diagnostic category (third row), we used the proportion of diagnostic rates in D2 and multiplied them with conversion rates within 1 year from ADNI1/GO/2 (see website FAQ). For estimating the average number of visits per subject (mean $\pm$ std.) in D4 (second row), we used the proportions for each diagnostic group and considered one visit per subject (ADNI procedures). We set the standard deviation to be zero, although in practice this won't be the case. 

For estimating the available biomarker data (lower half of table), we used a 1-year time-frame from start of ADNI2 (July 2012 -- July 2013) and computed the proportion of available data in that time frame. For AV1451, we used the same estimate as for AV45, due to the fact that the scan was introduced later on in ADNI2, and we expect more subjects to undergo AV1451 scans in ADNI3. A Python script that computes all the data from Table \ref{tab:biomk_data_available} is given in the TADPOLE repository: \url{https://github.com/noxtoby/TADPOLE/blob/master/statistics/tadpoleStats.py}.





% \bibitem[Prince(2014)]{prince2014world}
% Prince, M. and Jackson, J., 2014. World Alzheimer Report 2009, Alzheimer's Disease International.
% \end{thebibliography}

%\section*{References}

\bibliographystyle{elsarticle-harv}
\bibliography{bibliography}

% \printbibliography
% 
% \xpatchbibmacro{date+extrayear}{%
%   \printtext[parens]%
% }{%
%   \setunit{\addperiod\space}%
%   \printtext%
% }{}{}
% 
% \printbibliography

\end{document}